\documentclass[11pt,usletter]{amsart}
\usepackage[english]{babel}
\usepackage{array}
\usepackage[latin1]{inputenc}
\usepackage{amsmath}
\usepackage{amssymb}
\newcommand{\Izero}{{\mathrm{O}{\kern -0.6em 0}}}
\newtheorem{thm}{Theorem}[section]

\newtheorem{lemme}[thm]{Lemma}
\newtheorem{corol}[thm]{Corollary}

\newcommand\definesymb[1]{%
\expandafter\newcommand\csname #1#1\endcsname{\mathbb{#1}}%
}
\definesymb{N}\definesymb{Z}\definesymb{Q}\definesymb{R}
\definesymb{C}\definesymb{F}\definesymb{D}\definesymb{K}
\definesymb{U}

\newcommand{\place}{\vrule height 12pt depth 12pt width 0pt}
\newcolumntype{C}{>{\place \displaystyle}c<{\place}}
\newcolumntype{R}{>{\place \displaystyle}r<{}}
\newcolumntype{L}{>{\displaystyle}l<{\place}}
\newcolumntype{G}{l<{\place}}

\newcommand{\mat}[1]{{#1}}

\renewcommand{\vec}[1]{{#1}}
\title{Decidable and undecidable problems\\ about quantum automata}

\author{Vincent D. Blondel}
\address{Vincent D. Blondel\\
Departement of Mathematical Engineering\\
Université Catholique de Louvain.}
\email{blondel@inma.ucl.ac.be}

\author{Emmanuel Jeandel}
\address{Emmanuel Jeandel\\
Laboratoire de l'Informatique du Parall\'elisme\\
Ecole Normale Sup\'erieure de Lyon.}
\email{Emmanuel.Jeandel@ens-lyon.fr}

\author{Pascal Koiran}
\address{Pascal Koiran\\
Laboratoire de l'Informatique du Parall\'elisme\\
Ecole Normale Sup\'erieure de Lyon.}
\email{Pascal.Koiran@ens-lyon.fr}

\author{Natacha~Portier}
\address{Natacha Portier\\
Laboratoire de l'Informatique du Parall\'elisme\\
Ecole Normale Sup\'erieure de Lyon.}
\email{Natacha.Portier@ens-lyon.fr}

\let\sec=\section

\begin{document}
\bibliographystyle{alpha}
\maketitle

\begin{abstract}
%PK
We study the following decision problem: is the language recognized by
a quantum finite automaton empty or non-empty?
We prove that this problem is decidable or undecidable depending
%We prove that languages recognized by quantum finite automata are
%decidable or undecidable depending
on whether recognition is
defined by strict or non-strict thresholds. This result is in
contrast with the corresponding situation for probabilistic finite
automata for which it is known that strict and non-strict
thresholds both lead to undecidable problems. %languages.
\end{abstract}

\section{Introduction}\label{pres}

In this paper, we provide decidability and undecidability proofs
for two problems associated with quantum finite automata. Quantum
finite  automata (QFA) were introduced by Moore and Crutchfield
\cite{moore}; they are to quantum computers what finite automata
are to Turing machines.   Quantum automata are also analogous to
the probabilistic finite automata introduced in the 1960s by
Rabin that accept words with a certain probability \cite{rab5}
(see also \cite{paz} for a book-length treatment).
%PK
%Rather than
%accepting or rejecting words,
A quantum automaton $A$ assigns real
values $\mathrm{Val}_A(w)$ to input words $w$ (see below for a
precise description of how these values are computed).
%PK
$\mathrm{Val}_A(w)$ can be interpreted as the probability
that on any given run
of $A$ on the input word $w$, $w$ is accepted by $A$.
Associated
to a real threshold $\lambda$,  the languages recognized by the
automata $A$ with non-strict and strict threshold $\lambda$ are
$$L_{\geq}=\{w : \mathrm{Val}_A(w) \geq \lambda\} \mbox{ and } L_{>}=\{w : \mathrm{Val}_A(w) > \lambda\}.$$
Many properties of these languages are known in the case of
probabilistic and quantum automata. For instance, it is known that
the class of languages recognized by quantum automata is strictly
contained in the class of languages recognized by probabilistic
finite automata \cite{brodsky99characterizations}. For
probabilistic automata it is also known that the problem of
determining if $L_{\geq}$ is empty and the problem of determining
if $L_{>}$ is empty are undecidable \cite{rab1}. This is true even
for automata of fixed dimensions \cite{blca}.\\

In this contribution, we consider the problem of determining for a
quantum automata $A$ and threshold $\lambda$ if there exists a word
$w$ for which $\mathrm{Val}_A(w) \geq \lambda$ and if there
exists a word $w$ for which $\mathrm{Val}_A(w) > \lambda$. We
prove in Theorem \ref{t1} that the first problem is undecidable
and in Theorem \ref{t3} that the second problem is decidable. For
quantum automata it thus makes a difference to consider strict or
non-strict thresholds. This result is in contrast with
probabilistic automata for which both problems are undecidable.

Similarly to the languages $L_{\geq}$ and $L_{>}$, one can define
the languages $L_{\leq}$ and $L_{<}$ and ask whether or not they
are empty (of course, emptiness of $L_{\leq}$ is equivalent to
$L_>$ being equal to $\Sigma^*$). These two problems are known
\cite{rab1} to be undecidable for probabilistic automata. For
quantum automata our decidability results do again differ
depending on whether we consider strict or non-strict
inequalities. Our results are summarized in Table \ref{ta1}.\\

\begin{table}
 \label{ta1}
\begin{tabular}{l|c|c|c|c}
 & $L_{\geq}=\emptyset$ & $L_{>}=\emptyset$ & $L_{\leq}=\emptyset$ &
 $L_{<}=\emptyset$\\
PFA & undecidable & undecidable & undecidable & undecidable\\
 QFA &  undecidable & decidable & undecidable & decidable\\
 \end{tabular}
\smallskip
 \caption{Decidable and undecidable problems for probabilistic and
quantum automata.}
\end{table}

Before we proceed with the proofs, we first define what we mean by
a quantum finite automaton. A number of different quantum automata
models have been proposed in the literature and not all models are
computationaly equivalent. For the ``measure-many'' model of
quantum automata introduced by Kondacs and
Watrous~\cite{kondacs97power} the four problems of Table~1 are
proved undecidable in ~\cite{Jeandel02}. The model we consider
here is the so-called Measure Once Quantum Finite Automaton
introduced by Moore and Crutchfield \cite{moore}. These automata
operate as follows. Let $\Sigma$ be a finite set of input letters
and let $\Sigma^*$ denote the set of finite input words (including
the empty word); typical elements of $\Sigma^*$ will be denoted $w
= w_1 \cdots w_{|w|}$ where $w_i \in \Sigma$ and $|w|$ denotes the
length of $w$.  The QFA $A$ is given by a finite set of $n$
states, $n\times n $ unitary transition matrices $X_{\alpha}$ (one
for each symbol $\alpha$ in $\Sigma$), a (row) vector of unit norm
$s$ (the initial configuration), and a $n \times n$ projection
matrix $P$. Given a word $w \in \Sigma^*$, the value of $w$,
denoted $\mathrm{Val}_{A} (w)$, is defined by
$$
\mathrm{Val}_{A} (w) = \| \vec{s} {X}_w {P} \|^2
$$

In this expression, $\|\cdot\|$ is the euclidean vector  norm and
we use the notation ${X}_w$ for the product ${X}_{w_1} \cdots
{X}_{w_{|w|}}$. For a vector ${v}$, the value $\| \vec{v} \mat{P}
\|^2$ is the probability for the quantum state ${v}$ to be
observed in acceptance space. The value $\mathrm{Val}_{A} (w)$ can
thus be interpreted as the probability of observing the quantum
state in acceptance space after having applied the operator
sequence $X_{w_1}$ to $X_{w_{|w|}}$ to the initial
quantum state $s$.\\

%PK

The rest of this paper is organized as follows. In Section 2, we
reduce Post's correspondence problem to the problem of determining
if a quantum automata has a word of value larger than a given
threshold. Post's correspondence problem is undecidable and this
therefore proves our first result. Our reduction uses an encoding
of words in three dimensional space. In Section 3, we  prove
decidability of the same problem for strict inequality. For the
proof we use the fact that any compact matrix group is algebraic
and the group we consider can be given an effective description.

\sec{Undecidability for non strict inequality}

We prove in this section that the problem of determining if a
quantum automata has a word of value larger than some threshold is
undecidable. The proof is by reduction from Post's correspondence
problem (PCP), a well-known undecidable problem. An instance of
Post's correspondence problem is given by a finite alphabet
$\Sigma$ and $k$ pairs of words $(u_i, v_i) \in \Sigma^* \times
\Sigma^*$ for $i=1, \ldots, k$. A solution to the correspondence
is any non-empty
 word $w=w_1 \cdots w_n$ over the alphabet $\{1, \ldots, k\}$ such that
$u_w = v_w$, where $u_w = u_{w_1} \ldots u_{w_n}$. This
correspondence problem is known to be undecidable: there is no
algorithm that decides if a given instance has a solution
\cite{Pos-variant-rup}. It is easy to see that the problem remains
undecidable when the alphabet $\Sigma$ contains only two letters.
The problem is also known to be undecidable for $k=7$ pairs
\cite{matiyasevich96decision} but is decidable for $k=2$ pairs;
the decidability of the cases $2 < k <7$ is yet unresolved. We are
now ready to state our first result.

\begin{thm}
\label{t1}
  There is no algorithm that decides for a given automaton $A$ if there exists a word $w$
  for which $\mathrm{Val}_A(w) \leq 0$, or if there exists one for which $\mathrm{Val}_A(w) \geq 1$. These problems remain
  undecidable
  even if the automaton is given by 7 matrices in dimension 6, or by 2
  matrices in dimension 42.
% {\tt VB:Do we want to say something about the decidable cases in low
% dimension?}
\end{thm}

Proof. We proceed by reduction from  PCP. For our reduction we
need to encode words by unitary matrices. %
% (%
% a variant of the original proof of Swierczkowski can be found in \cite{su}
% EJ : Non. i. Ce n'est pas une variante, elle est profondément différente
% ii. C'est uniquement un cas très particulier, proche du notre..
% En conséquence, je déplace la référence plus loin
% )
We will take matrices that represent rotations of angle
$\arccos(3/5)$ on, respectively, the first and the third axis:

\[
\begin{array}{C C}
  \mat{X}_a = \frac{1}{5}\begin{pmatrix}
      3 &-4& 0 \\
      4 & 3& 0 \\
      0   & 0  & 5
 \end{pmatrix} &
\mat{X}_b = \frac{1}{5} \begin{pmatrix}
      5 & 0 & 0 \\
      0 & 3 &-4 \\
      0 & 4 & 3 \\
      \end{pmatrix}
\end{array}
\]

These matrices are unitary, $X_aX_a^T=I=X_b X_b^T$ and they
generate a free group since a result from Swierczkowski ensures us
that two irrationals rotations on orthogonal axes in $\RR^3$
generate a free group.
 %
% EJ
In addition to that, we now prove that there exists a vector $t$
such that $t X_u= t X_w$ implies $u = w$.
%{\tt VB. Désolé, je ne comprends pas cet argument. (Il
%me semble que ce dont nous avons besoin est: "$v X_u= v X_w$
%implies $u=w$).}

% EJ : voilà un échantillon de preuve
% tout mettre rendrait le discours beaucoup trop lourd pour un point
% qui n'est pas en soi le plus crucial (et dont la démonstration dans Su est
% tellement similaire (il n'y a que le choix du vecteur initial qui change)
% qu'on pourrait presque la recopier)

We will use here a method from~\cite{su}. One can show by
induction that for any reduced matrix product $M$ of $k$
matrices\footnote{A product is said to be {\em reduced} if no two
consecutive matrices in the product are inverse from each other.}
 taken from the set $\{X_a,
X_b, X_a^{-1}, X_b^{-1}\}$,  we have
$$(3\ 0\ 4) M = (x_1\ x_2\ x_3) / 5^k$$
with $x_1, x_2, x_3 \in \ZZ$, and
$5$ divides $x_2$ if and only if
% EJ Remarque : les précisions i. reduced matrix ii. k = 0 (ligne suivante)
% sont /indispensables/.  Sans la précision k = 0, le résultat est non
% seulement faux, mais aussi inutilisable (y réfléchire) dans la suite
$k=0$ (and then $M=I$).

 The result is obviously true for $k = 0,1$. Now, if
$M = M' X_1 X_0$, then
$(3\ 0\ 4) M = (x_1\ x_2\ x_3) /5^k X_a X_b = ( x_4\ x_5\ 5 x_3) / 5^{k+1} X_b$
for some $x_4, x_5$, and by induction hypothesis $5$ does not divide $x_5$.
Now $(3\ 0\ 4) M = (x_6\ 3 x_5 +  20 x_3\ x_7) / 5^{k+2}$ so that $5$ does
not divide the second term. The proof for all the other cases is similar.

We will now call $\vec{t}$ the row vector  $(3\ 0\ 4)$. If
$\vec{t} X_u  = \vec{t} X_v$ then  $\vec{t} X_u X_v^{-1} =
\vec{t}$. As the second component of $\vec{t}$ is equal to $0$,
the product must be trivial, and so $u=v$.

Given an instance
%PK
$(u_i, v_i)_{1 \leq i \leq k}$ of PCP over the alphabet $\{a,b\}$
and a word $w \in \{1, \ldots, k\}^*$, we construct  the matrix

\[ Y_w =
     \frac{1}{2}
 \begin{pmatrix}
    \mat{X}_{u_w} + \mat{X}_{v_w}& \mat{X}_{u_w} - \mat{X}_{v_w} \\
    \mat{X}_{u_w} - \mat{X}_{v_w}& \mat{X}_{v_w} + \mat{X}_{u_w} \\
\end{pmatrix}
\]
These matrices are unitary, and verify $\mat{Y}_{w \nu} =
\mat{Y}_w \mat{Y}_\nu$

A solution of the original PCP problem is a nonempty word $w \in
\{1, \ldots, k\}^*$ such that the upper-right block of the matrix
$\mat{Y}_w$ is equal to zero. We may use the previously introduced
vector $\vec{t}=(3\ 0\ 4)$ to test this condition. We have
$$
\begin{pmatrix}
   \vec{t} & \vec{0}
\end{pmatrix} \mat{Y}_w
=
%PK
{1 \over 2}
\begin{pmatrix}
    \vec{t}\mat{X}_{u_w} +\vec{t} \mat{X}_{v_w} &
    \vec{t}\mat{X}_{u_w} - \vec{t}\mat{X}_{v_w}
\end{pmatrix}
$$
 and thus a solution of the PCP problem is a word $w$ such that the
%PK third
last three coordinates of $\vec{y}\mat{Y}_w $ are equal to zero,
where $\vec{y} = \begin{pmatrix} \vec{t} & \vec{0}
\end{pmatrix}$.
 This condition can be tested with
  a projection matrix. Defining
\[
  \mat{P} = \begin{pmatrix}
    0_3 & 0 \\
    0 & I_3 \\
    \end{pmatrix}
  \]
we have that the solutions of the original PCP problem are the
words $w$ for which $\vec{y} \ \mat{Y}_w \ \mat{P} = 0$,
which is equivalent to
  $$\mathrm{Val}_A(w)=\| \vec{y}
\mat{Y}_w \mat{P}
   \|^2 = 0$$
 The
values taken by $\mathrm{Val}_{A} (w)$ are non-negative and so the
problem  of determining if there exists a word $w$ such that
$\mathrm{Val}_{A} (w) \leq 0$ is undecidable. Notice also that
$\|
\vec{y}
\mat{Y}_w \mat{I}
   \|^2 = 1$ and so
  $$ \| \vec{y}
\mat{Y}_w \mat{(I-P)}
   \|^2 \leq 1$$
   with equality only for $\vec{y}
\mat{Y}_w \mat{P}
 = 0.$
   Thus, the problem of determining if there exists a word $w$ such that
$\mathrm{Val}_{A} (w) \geq 1$ is undecidable.

We now show how to reduce the number of matrices to two. We use a
construction from Blondel \cite{blondel97,blca}. Given the above
matrices $\mat{Y}_i$ and the projection matrix $P$,
%PK and the vector $y=(v 0)$,
% \vec{y} est maintenant défini plus haut
we define
\[
  \mat{Z}_0 = \begin{pmatrix}
    \mat{Y}_1 & 0 & \hdots & 0 \\
    0 & \mat{Y}_2 & \ddots & 0 \\
    \vdots & \vdots & \ddots & \vdots \\
    0 & 0 & \hdots & \mat{Y}_7 \\
\end{pmatrix}
\mbox{ and } \mat{Z}_1 = \begin{pmatrix}
  0 &I & 0 & 0 \\
  0 & \ddots & \ddots & 0 \\
  \vdots & \vdots & \ddots& I \\
  I & 0 & \hdots & 0 \\
  \end{pmatrix}
  \]

When taking products of these two matrices the matrix $\mat{Z_1}$
acts as a ``selecting matrix" on the blocks of $Z_0$. Let us
define $\vec{x} =
\begin{pmatrix} \vec{y} & 0
\end{pmatrix} $ and
% EJ Cette matrice ne me plaisait pas, puisqu'on ne voit que peu sa dimension
% \[\mat{Q} = \begin{pmatrix} \mat{P} & 0 \\ 0 & 0  \end{pmatrix}\]
\[  \mat{Q} = \begin{pmatrix}
    \mat{P} & 0 & \hdots & 0 \\
    0 & 0  & \ddots & 0 \\
    \vdots & \vdots & \ddots & \vdots \\
    0 & 0 & \hdots & 0 \\
\end{pmatrix}
\]

We claim that there exists a non-empty word $w$ over the alphabet
$\{1, \dots, 7\}$ such that ${\| \vec{y} \mat{Y}_w \mat{P} \| =
0}$ if and only if there exists a non-empty word $\nu$ over
$\{0,1\}$ such that $\| \vec{x} \mat{Z}_\nu \mat{Q} \| = 0$. The
complete proof of this claim is given in \cite{blondel97} and
is not reproduced here. \hfill $\square$\\

It is possible to give a stronger form to the second part of the
Theorem. We prove below that, whatever threshold $0 \leq \lambda <
1$ is used, the problem of determining if there exists a word for
which $\mathrm{Val}_A(w) \geq \lambda$ is undecidable. This result
follows as a corollary of the following lemma.

\begin{lemme}
\label{l1}
  Associated to every QFA $\mathcal{A}$ and threshold $\lambda < 1$
  we can effectively construct a QFA ${B}$ such that the language
  recognized with
  threshold $\lambda$ by ${B}$ is the language recognized with
  threshold $0$ by ${A}$. Moreover, if $\lambda \in \QQ$ and $A$ has
  only rational entries then $B$ can be chosen with rational entries.
\end{lemme}

Proof. The idea is to construct $B$ by adding a state to $A$. Let
$A$ be given by the unitary matrices $X_i^A$, the projection
matrix $P^A$ and the initial vector $s^A$. Let
\[ \mat{X}_i^{B} =
  \begin{pmatrix}
    \mat{X}_i^{A} & 0 \\
    0 & 1 \\
  \end{pmatrix}
  \]
and define $s^B=\begin{pmatrix}
  \sqrt{\lambda} \, \vec{s}^A &\sqrt{1 - \lambda}
\end{pmatrix}$. If we choose \[
  \mat{P}^{B} =
  \begin{pmatrix}
    \mat{P}^{A} & 0 \\
    0 & 0 \\
  \end{pmatrix}
\]
we immediately have $\mathrm{Val}_{B} (w) = \lambda \;
\mathrm{Val}_{A}(w)$ and the first part of lemma is proven. The
entries $\sqrt{\lambda}$ and $\sqrt{1 -\lambda}$ do in general not
need to be rational. It remains to show how the parameters of $B$
can be chosen rational when those of $A$ are. We therefore use
Lagrange's theorem to write $\lambda$ and $1-\lambda$ as the sum
of the square of four rational numbers, say
$\lambda=a_1^2+a_2^2+a_3^2+a_4^2$ and
$1-\lambda=b_1^2+b_2^2+b_3^2+b_4^2$.

 Now, if
we define
\[ \vec{s^B} =
\begin{pmatrix}
  a_1 \vec{s^A} & a_2 \cdots a_4& b_1 \cdots b_4
\end{pmatrix} \;
\mat{X}^B_i = \begin{pmatrix}
  X_i^A & 0 \\
  0  & I_7 \\
  \end{pmatrix}
  \;
\mat{P^B} = \begin{pmatrix}
  P^A & 0 & 0 \\
  0 & I_3 & 0\\
  0 & 0 &  0_4 \\
\end{pmatrix} \]
we have immediately $\mathrm{Val}_{B} (w) = a_1^2
\mathrm{Val}_{A}(w) + a_2^2 + a_3^2 + a_4^2$,
$\| \vec{s}^B \| ^2 =1$
and the lemma is proved.
\hfill $\square$\\

Combining Lemma \ref{l1} with Theorem \ref{t1}, we immediately
obtain:

\begin{corol}
  For any rational $\lambda$, $0 \leq \lambda < 1$, there is no
 algorithm that
 decides if a given quantum automata has a word $w$ for which
 $\mathrm{Val}(w) \leq \lambda$.
\end{corol}

\sec{Decidability for strict inequality}

We now prove that the problem of determining if a quantum automata
has a word of value \emph{strictly}  larger than some threshold is
decidable. This result points to  a difference between quantum and
probabilistic automata since for probabilistic automata this
problem is known to be undecidable.

Once an automaton is given one can of course always enumerate all
possible words $w$ and halt as soon as one is found for which
$\mathrm{Val}_{A}(w) > \lambda$, and so the problem is clearly
semi-decidable. In order to show that it is decidable it remains
to exhibit a procedure that halts when $\mathrm{Val}_{A}(w) \leq
\lambda$ for all $w$.

Let a quantum automata $A$ be given by a finite set of $n\times n
$ unitary transition matrices $X_i$, an initial configuration $s$
of unit norm, and a projection matrix $P$. The value of the word
$w$ is given by $ \mathrm{Val}_{A} (w) = \| \vec{s} {X}_w {P}
\|^2$. Let $\mathcal{X}$ be the semigroup generated by the
matrices $X_i$, $\mathcal{X}=\{X_w: w \in \Sigma^*\}$, and let
$f:\RR^{n\times n} \mapsto \RR$ be the function defined by
$f(X)=\|s {X} P \|^2$. We have that
$$\mathrm{Val}_{A}(w)=f(X_w)$$ and the problem is now that of
determining if $f(X) \leq \lambda$ for all $X \in \mathcal{X}$.
The function $f$ is a (continuous) polynomial map and so this
condition is equivalent to $f({X}) \leq \lambda$ for all $X \in
\overline{\mathcal{X}}$, where $\overline{\mathcal{X}}$ is the
closure of $\mathcal{X}$ in $\RR^{n \times n}$. The set
$\overline{\mathcal{X}}$ has the interesting property that  it is
algebraic (see below for a proof), and so there exist polynomials
mappings $f_1,  \ldots, f_p: \RR^{n \times n} \mapsto \RR$, such
that $\overline{\mathcal{X}}$  is exactly the set of common zeroes
of $f_1, \ldots, f_p$. If the polynomials $f_1,  \ldots, f_p$ are
known, the problem of determining whether $f(\mat{X}) \leq
\lambda$ for all ${X} \in \overline{\mathcal{X}}$ can be written
as a  quantifier elimination problem
\begin{equation} \label{eq}
 \forall X \big[ \left(f_1(X) = 0 \wedge \dots \wedge f_p(X) =
0\right) \implies f(X) \geq \lambda \big].
\end{equation}
This is a first-order
formula over the reals, and can be decided effectively by
Tarski-Seidenberg elimination methods (see
\cite{renegar92computational} for a survey of known algorithms).
%PK
If we knew how to effectively compute the polynomials
$f_1,  \ldots, f_p$ from the matrices $X_i$, a decision algorithm
would therefore follow immediately.
In the following we solve a simpler problem:
we effectively compute a sequence of polynomials whose zeroes
describe the same set $\overline{\mathcal{X}}$ after finitely many
terms (but we may never know how many).
It turns out that this is sufficient for our purposes.

\begin{thm}
\label{t3}
%PK
Let $(X_i)_{i \in \Sigma}$ be unitary matrices of dimension $n$ and let
$\overline{\mathcal{X}}$ be the closure of the semigroup
%PK
$\{X_w: w \in \Sigma^*\}$.
The set $\overline{\mathcal{X}}$ is algebraic,
%PK
and if the $X_i$ have rational entries we can
effectively compute a sequence of polynomials $f_1,  \dots, f_i,
\dots$ such that
  \begin{enumerate}
%PK
    \item If $X  \in \overline{\mathcal{X}}$,
$f_i (\mat{X}) = 0$ for all $i$;
    \item There exists some $k$ such that $\overline{\mathcal{X}}=\{ X : f_i(X)=0, i=1, \ldots, k\}$.
  \end{enumerate}
\end{thm}

Proof. We first prove that $\overline{\mathcal{X}}$ is algebraic.
It is known (see, e.g., \cite{onishchik90lie}) that every compact
group of real matrices is algebraic.
%of matrices over $\RR$ is a Lie group over $\RR$ and is
%therefore algebraic.
In fact, the proof of algebraicity in~\cite{onishchik90lie} reveals
that any compact group $G$ of real matrices of size $n$ is the zero
set of
\[
  \RR[X]^G = \{ f \in \RR[X] : f(I) = 0 \mbox{ and } f(g X) = f(X)
  \mbox{ for all } g \mbox{ in } G\},
\]
i.e., $G$ is the zero set of the polynomials in $n \times n$ variables
which vanish at the identity and are invariant under the action of
$G$. We will use this property later in the proof.

To show that $\overline{\mathcal{X}}$ is algebraic,
it suffices to show that
$\overline{\mathcal{X}}$ is compact and is a group. The set
$\overline{\mathcal{X}}$ is obviously compact (bounded and closed
in a normed vector space of finite dimension). Let us show that it
is a group.
%PK The quickest way to see that it is a group is to say
%that every compact subsemigroup of a topological group is in fact
%a subgroup.
It is in fact known that every compact subsemigroup of a topological
group is a subgroup. Here is a self-contained proof in our setting:
for every
matrix $\mat{X}$, the sequence $\mat{X}^k$ admits a subsequence
that is a Cauchy sequence, by compactness. Hence for every $\epsilon$
there exists $k > 0$ and $l > k + 1 $ such that $\| \mat{X}^k -
\mat{X}^l \| \leq \epsilon$, that is $\| \mat{X}^{-1} -
\mat{X}^{l-k-1} \| \leq \epsilon$ (recall that $\|\mat{A} \mat{B}
\| = \| \mat{B} \|$ if $\mat{A}$ is
%PK orthonormal
unitary, and if $\| . \|$ is the operator norm associated to the
Euclidean norm). Hence,
$\mat{X}^{-1}$ is in the set and the first part of the theorem is
proved.
For notational convenience, we will denote the group
$\overline{\mathcal{X}}$ by $G$ in the remainder of the proof.

For the second part of the theorem, we will prove that we can take
\[
  \{f_i\} = \{f \in \QQ[X] :  f(I) = 0 \mbox{ and } f(\mat{X}_jX) =
  f(X) \mbox{ for all } j \mbox{ in } \Sigma\}
  \]
%PK J'ai fait de nombreux changements dans la suite de l'article.
%En particulier, j'ai enlevé à contrecoeur la remarque historique sur
%les algèbres parce qu'on utilise plus de théorème sur les algèbres!
In words, this is the  set $\QQ[X]^G$ of rational polynomials
which vanish at the identity and
  are invariant under the action of each matrix $\mat{X}_j$.
It is clear that this set is recursively enumerable.
We claim that $G$ is the zero set of the $f_i$'s.
By Noetherianity the zero set of the $f_i$'s is equal to the zero set
  of a finite subset of the $f_i$'s, so that
the theorem follows immediately from this claim.
%For notational convenience, let us denote the group
%$\overline{\mathcal{X}}$ by $G$ and
To prove the claim, we will use the fact that $G$
is the zero set of $\RR[X]^G$.
%\[
%  \RR[X]^G = \{ f \in \RR[X] : f(I) = 0 \mbox{ and } f(g X) = f(X)
%  \mbox{ for all } g \mbox{ in } G\}.
%\]
%It is known that $G$ is the zero set of $\RR[X]^G$~\cite{onishchik90lie}.
Note that \[
  \RR[X]^G = \{ f \in \RR[X] : f(I) = 0 \mbox{ and }
f(X_j X) = f(X) \mbox{ for all } j \mbox{ in } \Sigma\}
\]
(a polynomial is invariant under the action of $G$ if and only it is
invariant under the action of all the $X_j$).
This observation implies immediately that each $f_i$ is in $\RR[X]^G$,
so that the zero set of the $f_i$'s contains the zero set of
$\RR[X]^G$. The converse inclusion follows from the fact that any
element $P$ of $\RR[X]^G$ can be written as a linear combination of some
$f_i$'s. Indeed, let $d$ be the degree of $P$ and let $E_d$ be the set of
real polynomials in $n \times n$ variables of degree at most $d$.
The set $V_d = E_d \cap \RR[X]^G$ is a linear subspace of $E_d$
defined by by a system of linear equations with rational coefficients
(those equations are $f(I) = 0$
and $f(X_j X) = f(X)$ for all $j \in \Sigma$).
Hence there exists a basis of $V_d$ made up of polynomials with
rational coefficients, that is, of elements of $\{f_i\}$.
This completes the proof of the claim, and of the theorem.
%By a theorem due to Hilbert, it is known that $G$
%is precisely the set of zeroes of $\RR[X]^G$, and that $\RR[X]^G$
%is finitely generated as an algebra\footnote{This result was
%proved by Hilbert; he proved the Nullstellensatz as a lemma for
%proving this result.}(\cite{sturm}). It is also known that the
%condition $\forall g \in G \big[ f(g X) = f(X) \big]$ is equivalent to
%$\forall i\in \Sigma \big[ f(X_i X) = f(X) \big]$.
%Since this last
%equation is linear in terms of  the coefficients of $f$ (this also
%holds for the equation $f(I) = 0$), generators of $\RR[X]^G$ can
%be taken to be in $\QQ[X]$.

%{\tt VB. Références pour ces résultats?}

%Summing up, we infer that $G$ is exactly the sets of zero of
%\[
%  \{f_i\} = \{f \in \QQ[X] :  f(I) = 0 \mbox{ and } f(X \mat{X}_i) = f(X), \f%orall i \in \Sigma\}
%  \]
%This set is effectively enumerable and is finitely generated, so
%it satisfies the conditions we need and the proof is complete.
\hfill $\square$\\

We may now apply this result to quantum automata.

\begin{thm} \label{deci}
The two following problems are decidable.
\begin{itemize}
\item[(i)] Given a quantum automaton $A$ and a threshold $\lambda$, decide
whether there exists a word $w$ such that
$\mathrm{Val}_{A}(w)>\lambda$.
\item[(ii)] Given a quantum automaton $A$ and a threshold $\lambda$, decide
whether there exists a word $w$ such that
$\mathrm{Val}_{A}(w) < \lambda$.
\end{itemize}
\end{thm}

Proof. We only show that problem (i) is decidable. The argument
for problem (ii) is essentially the same.

As pointed out at the beginning of this section, it suffices to
exhibit an algorithm which halts if and only if $Val_A(w) \leq \lambda$ for
every word $w$.
%It follows from Theorem~\ref{t3} that the following
%algorithm has the required property:
%PK
Consider the following algorithm:
\begin{itemize}
\item enumerate the $f_i$'s;
\item for every initial segment $f_1,\ldots,f_p$, decide
whether~(\ref{eq}) holds, and halt if it does.
\end{itemize}
It follows from property~(1) in Theorem~\ref{t3} that $Val_A(w) \leq
\lambda$ for every word $w$ if the algorithm halts. The converse
follows from property~(2).
\hfill $\square$\\

Throughout the paper we have assumed that our unitary matrices have
rational entries.
It is not hard to relax this hypothesis. For instance, it is clear
from the proofs that Theorems~\ref{t3} and~\ref{deci} can be
generalized to matrices with real algebraic entries.

In the proof of Theorem~\ref{deci} we have bypassed the problem of
explicitly computing a finite set of polynomials defining
$\overline{\mathcal{X}}$. It is in fact possible to show that this
problem is algorithmically solvable~\cite{DJK03}. This implies in
particular that the following two problems are decidable:
\begin{itemize}
\item[(i)] Decide whether a given treshold is isolated.
\item[(ii)] Decide whether a given QFA has an isolated threshold.
\end{itemize}
A
threshold $\lambda$ is said to be isolated if:
\[
  \exists \epsilon > 0 \; \forall w \in \Sigma^*\;
  | \mathrm{Val}_{A}(w) - \lambda | > \epsilon.
  \]
It is known that these two problems are undecidable for probabilistic
automata~\cite{Bertoni75,BMT77,blca}.

%Actually there is a more straightforward procedure as we are able to
%find the polynomials $f_1, \dots, f_k$ exactly  (
%{\tt NP : mettre la ref du truc en cours})
%The decision procedure became: compute
%$f_1, \dots, f_k$ and apply quantifier elimination.

%Computing $f_1 \dots f_k$ is also be helpful to prove
%the decidability of problems related to isolated thresholds. A
%threshold $\lambda$ is said to be isolated if:
%\[
%  \exists \epsilon > 0 \; \forall w \;
%  | \mathrm{Val}_{A}(w) - \lambda | > \epsilon
%  \]
%This condition
%  is equivalent to the following first order predicate
%  \[
%    \exists \epsilon > 0\; \forall \mat{X} \: \left( f_1(\mat{X}) = \dots =
%f_p(\mat{X}) = 0 \implies  | f(\mat{X})  - \lambda | >
%\epsilon\right)
%\]
%and the problem of determining if a threshold is isolated would
%therefore be decidable if an effective procedure was available for
%finding the polynomials $f_1, \dots, f_k$. Notice that the
%polynomial sequence used in Theorem \ref{t3} do in this case not
%suffice for deriving a decision procedure. The same comments apply
%to the problem of deciding if an automata has an isolated threshold
%and these problems are both undecidable for probabilistic finite
%automata.

The algorithm of ~\cite{DJK03} for computing
$\overline{\mathcal{X}}$ has also applications to quantum
circuits: this algorithm can be used to decide whether a given set
of quantum gates is complete ({\em complete} means that any
unitary transformation can be approximated to any desired accuracy
by a quantum circuit made up of gates from the set). Much effort
has been devoted to the construction of specific complete sets of
gates~\cite{deutsch95universality,barenco95elementary}, but no
general algorithm for deciding whether a given set is complete was
known.

Finally, we note that the proof of Theorem~\ref{deci} does not yield
any bound on the complexity of problems (i) and (ii). We hope to
investigate this question in future work.

{\small \section*{Acknowledgment}
P.K. would like to thank Etienne Ghys for pointing out reference~\cite{onishchik90lie}.}

%\bibliography{qfa}

\newcommand{\etalchar}[1]{$^{#1}$}

\end{document}